\documentclass[a4paper,aps,prl,twocolumn,superscriptaddress,floatfix,showpacs]{revtex4}
\usepackage{amsfonts}
\usepackage{amssymb}
\usepackage{graphicx}
\usepackage{texdraw}
\usepackage{color}

\begin{document}

\title{Experimental observation of quantum entanglement in low dimensional
spin systems}
\author{T. G. Rappoport, L. Ghivelder}
\affiliation{Instituto de F\'{\i}sica, Universidade Federal do Rio de Janeiro, Caixa
Postal 68.528-970, Rio de Janeiro, Brazil}
\author{J. C. Fernandes, R. B. Guimar\~{a}es, M. A. Continentino}
\affiliation{Instituto de F\'{\i}sica, Universidade Federal Fluminense, Campus da Praia
Vermelha, Niter\'{o}i, 24210-340, Brazil}
\date{\today}

\begin{abstract}
We report macroscopic magnetic measurements carried out in order to detect
and characterize field-induced quantum entanglement in low dimensional
spin systems. We analyze the pyroborate $\mbox{MgMnB$_2$O$_5$}$ and the and the warwickite
$\mbox{MgTiOBO$_3$}$, systems with spin 5/2 and 1/2 respectively. By using the magnetic
susceptibility as an entanglement witness we are able to quantify
entanglement as a function of temperature and magnetic field. In addition,
we  experimentally distinguish for the first time a random singlet phase from a Griffiths
phase. This analysis opens the possibility of a more detailed characterization of low dimensional materials.

\end{abstract}

\maketitle



%%%%%%%%%%%%%%%%%%%%%%%%%%%%%%%%%%%%%%%%%%%%%%%%%
%%%%%%%%%%%%%%%%%%%% Introduction %%%%%%%%%%%%%%%%%%%%%%%%

Since the development of quantum mechanics, entanglement has been a subject
of great interest. Lately, this is mainly due to the importance of
entanglement in quantum information and computation. As a consequence, a
great effort has been made to detect and quantify entanglement in quantum
systems~\cite{optics}. In addition, quantum spin chains, a class of systems
well known in solid state physics, began to be studied in the framework of
quantum information theory; there are proposals for the use of such systems
in quantum computation~\cite{qc}. Naturally, entanglement in interacting
spin chains acquired relevance in the QI community. Therefore, there has
been a special effort in understanding and quantifying quantum entanglement
in solid-state~\cite{lidar,vidal,vedral,vedral1}. At the same time, the
condensed matter community has begun to notice that entanglement may play a
crucial role in the properties of different materials~\cite{aeppli}.

It is a difficult task to determine experimentally if a state is entangled
or not. One of the new promising methods for entanglement detection is the
use of an entanglement witnesses (EW). EW are observables which have
negative expectation values for entangled states. Magnetic susceptibility
was recently proposed as an EW~\cite{vedral} and some old experimental
results were re-analyzed wthin this new framework~\cite{sus-ew-exp}.

It has been known for a long time that entanglement appears in quantum spin
chains, like the spin $1/2$ Heisenberg model. The disordered spin $1/2$
one-dimensional Heisenberg model, for example, presents a random singlet
phase (RSP), where singlets of pairs of arbitrarily distant spins are formed~\cite{mdh}. 
Although entanglement was already known to exist in such chains,
it had not been quantified theoretically until this decade~\cite{refael04}.
A previous study of a diluted magnetic material~\cite{aeppli} has shown the
importance of entanglement, but to our knowledge, this is the first
experimental measurement of quantum entanglement in a magnetic material. As
representative systems, we analyze the pyroborate MgMnB$_{2}$O$_{5}$~\cite%
{F1,F2} and the warwickite MgTiOBO$_{3}$~\cite{F3}, two quasi-one
dimensional disordered spin compounds with previously known magnetic and
thermodynamic properties that suggest the existence of either a RSP
or a Griffiths phase (GP)~\cite{griffiths} at low temperatures.

There are no experimental studies on random magnetic chains which
discriminate these two phases. In this Letter, from a detailed analysis of
magnetic measurements, we show unambiguously the existence of a RSP in $%
\mbox{MgTiOBO$_3$}$, which is expected for a spin-1/2 random exchange
Heisenberg antiferromagnetic chain (REHAC). In addition, our study of $%
\mbox{MgMnB$_2$O$_5$}$\ provides experimental evidence for the existence of
a Griffiths phase in a low dimensional system with $S>1/2$.

Addressing the entanglement properties of these random systems, there is
also a clear distinction between the RSP and the Griffiths phase. For the
former, entanglement is well characterized and has been shown to scale with
the logarithm of the size~\cite{refael04,laflorencie}. For the latter there
is no theoretical study of how entanglement behaves.

We make use of magnetization and ac susceptibility measurements as a
function of temperature and applied magnetic field to detect and quantify
entanglement by using the susceptibility as an entanglement witness~\cite%
{vedral}. First we show that both systems present entanglement at low
temperatures with no applied field. Next, we analyze the ac. susceptibility
and magnetization as function of field for different temperatures and we
quantify the variation of entanglement as a function of applied field. We
observe that entanglement increases for increasing magnetic fields in a
region of the B$\times $T diagram. This unusual behavior was suggested by
Arnesen \textit{et al.} ~\cite{vedral1} and called magnetic entanglement.

In both pyroborate MgMnB$_{2}$O$_{5}$ and warwickite MgTiOBO$_{3}$ there are
ribbons formed by oxygen octahedra sharing edges. These octahedra give rise
to four columns, along the ribbons, whose centers define a triangular
lattice and two different crystallographic sites for metals: one in the
central columns and another in the border ones. In the pyroborate such
columns do not touch and both metal sites are equally occupied by the two
metal ions \cite{UB}. In the warwickite, on the contrary, the columns do
touch and the metal sites are probably occupied as in MgScOBO$_{3}$: 76 $\%$
of the internal sites occupied by the transition metal and 24 $\%$ by the
alkaline-earth metal. The sites on the border columns have the opposite
occupancy \cite{Norr}.

The pyroborate powder was obtained from grinded single crystals, and the
warwickite powder was directly obtained from its synthesis. The warwickite
sample was analyzed through X-Ray diffractometry; it has been verified that
the material was well crystallized and that its purity was 97.72 $\%$ , as
evaluated by the method of Lutterotti et al. \cite{Lu}. The more abundant
impurity was the non-magnetic MgTiO$_{5}$. More details on sample
preparation were previously published~\cite{F2,F3}. Dc magnetization and ac
susceptibility measurements were performed with a commercial apparatus
(Quantum Design PPMS).

In $\mbox{MgTiOBO$_3$}$, evidence for a RSP-like behaviour was previously
obtained from specific heat $C(T)$, susceptibility $\chi (T)$ and
magnetization $m(H)$ measurements ~\cite{F3}. These quantities exhibit the
characteristic power law behavior associated with a RSP, $\chi (T)\propto
T^{-\alpha }$, down to the lowest measured temperature. In this system the
magnetic ion $Ti^{3+}$ has spin S=1/2, and due to the negligible magnetic
anisotropy this material is well described by a spin-1/2 (REHAC). The
physical behavior is controlled by an infinite randomness fixed point
independent of the amount of disorder. On the other hand in the $%
\mbox{MgMnB$_2$O$_5$}$\ pyroborate, the magnetic ion $Mn^{2+}$ is a spin
5/2, $S$ state ion. The phase diagram of a REHAC with $S\geq 1/2$ is not a
trivial one. For weak disorder these systems present GP, while only for
strong disorder a RSP appears~\cite{saguia}.

\begin{figure}[t]
\includegraphics[width=0.9\columnwidth,clip]{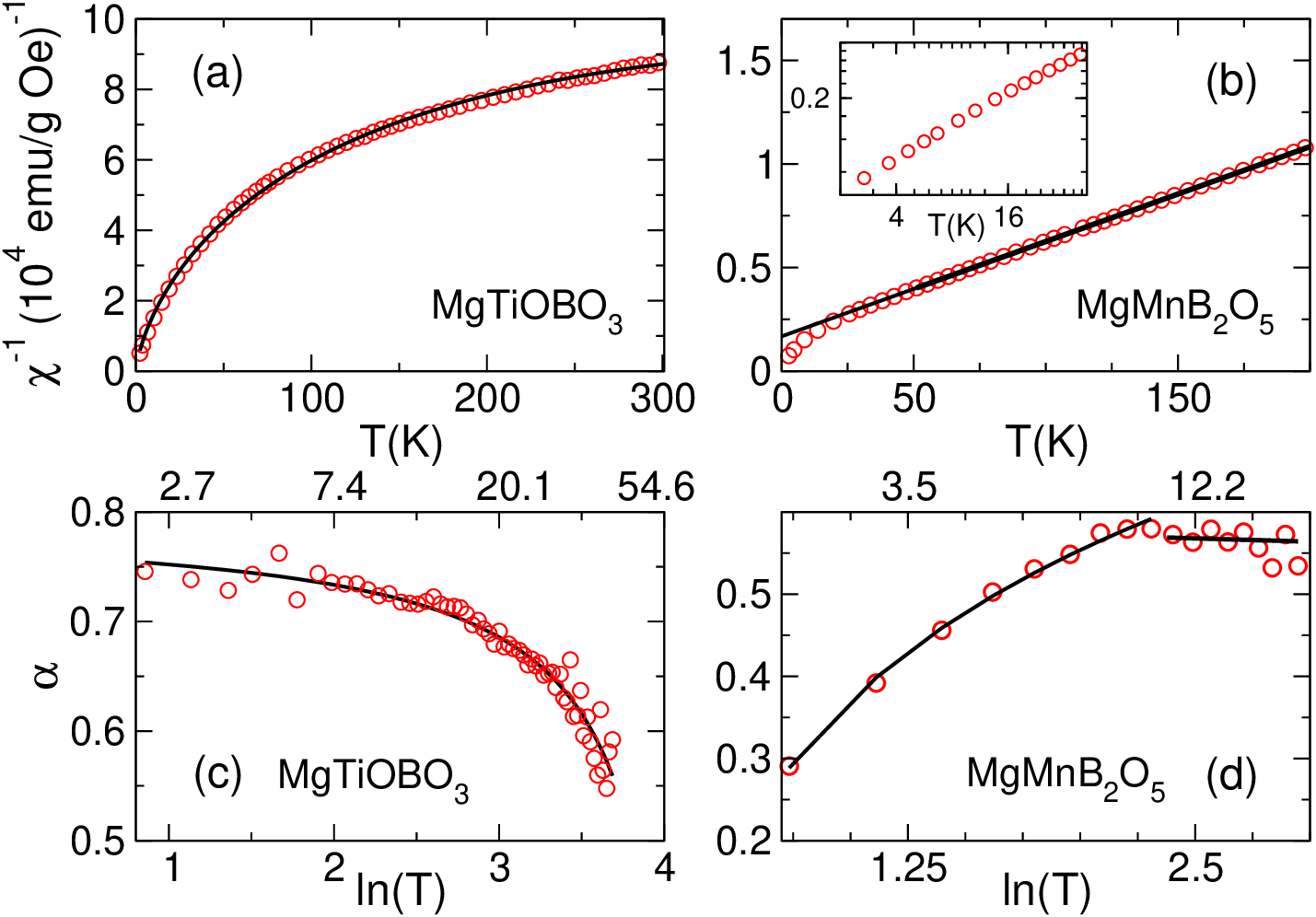}
\caption{Up: Magnetic susceptibility versus temperature. (a) Experimental
data for $\mbox{MgTiOBO$_3$}$\ (open circles) and fitting using the
susceptibility expression of a RSP (solid line). On panel (b), the
experimental data for $\mbox{MgTiOBO$_3$}$\ (open circles) and for high
temperatures a Curie-Weiss fiiting (solid line). Down: Exponent $\protect%
\alpha $ of $\protect\chi \propto T^{\protect\alpha }$ as a function of $\ln
T$ for (c) $\mbox{MgTiOBO$_3$}$\ and (d) $\mbox{MgMnB$_2$O$_5$}$\ . }
\label{Fig1}
\end{figure}

In figure \ref{Fig1}, panels (a) and (b), we show the ac magnetic
susceptibility as a function of temperature for $\mbox{MgTiOBO$_3$}$\ and $%
\mbox{MgMnB$_2$O$_5$}$\ respectively. Both systems have a sub-Curie regime
at low temperatures. $\mbox{MgMnB$_2$O$_5$}$\ acquires a Curie-like
temperature dependence around 50 K. On the other hand, $\mbox{MgTiOBO$_3$}$\
presents a sub-Curie susceptibility even at room temperature. It is known
that both systems have a susceptibility which behaves as $\chi (T)\propto
T^{-\alpha }$, although the temperature dependence of $%
\alpha $ was not further analyzed.

These two different phases should be distinguished experimentally by the
temperature dependence of the exponent $\alpha $. The GP is characterized by a 
constant value of $\alpha $. For the RSP, we should expect a low-temperature 
susceptibility following $\chi (T)=\frac{1}{T\ln ^{2}(\Omega _{0}/T)}$~\cite{fisher94},
which is equivalent to $\alpha (T)=1-\frac{a}{\ln (T/\Omega _{0})}$, where $%
a $ is a constant~\cite{hirsch}. So, the RSP is characterized by a slowly
varying $\alpha (T)$.

Following Hirsch~\cite{hirsch}, we analyze the data by redefining $\alpha
=-d(\ln (\chi ))/d(\ln (T))$ and extract the temperature dependence of the
exponent $\alpha (T)$ for both samples. Furthermore, we fit the experimental
data of Fig \ref{Fig1}(a), using $1/\chi (T)=T\ln ^{2}(\Omega _{0}/T)$
(solid line). Both Figs.~\ref{Fig1}(a) and (c) indicate that the
susceptibility coincides exactly with the RSP model and $\alpha $ follows
the same tendency previously predicted by numerical calculations~\cite%
{hirsch}. In $\mbox{MgMnB$_2$O$_5$}$, previous assessments and the inset of
Fig.~\ref{Fig1}(b) suggest a power law behavior for $\chi (T)$ with a
constant $\alpha \sim 0.55$. Within a more detailed analysis, shown in Fig.~%
\ref{Fig1}(d) we see an unequivocal slow increase of the exponent $\alpha $,
followed by a constant regime at intermediate temperatures. Although $\alpha 
$ is not constant in the whole temperature interval, as expected for a
Griffiths phase (GP), its increase with T is clearly inconsistent with a
phase governed by an infinite randomness fixed point or RSP. However, for
temperatures higher than 7 K, $\alpha $ is constant (up to 20 K), and this
strongly supports the existence of a GP in this system. In fact the
variation of $\alpha $ at low T may be related to a freezing of the Mn
moments, as suggested by a low temperature anomaly in the specific heat of
this material \cite{F1}. %%%%%%%%%%%%%%%%%%%%%%%%%%%%%%%%%%%%%%%%%%%%%%%%
%%%%%%%%%%%%%%%%%%

\begin{figure}[t]
\includegraphics[width=1.0\columnwidth,clip]{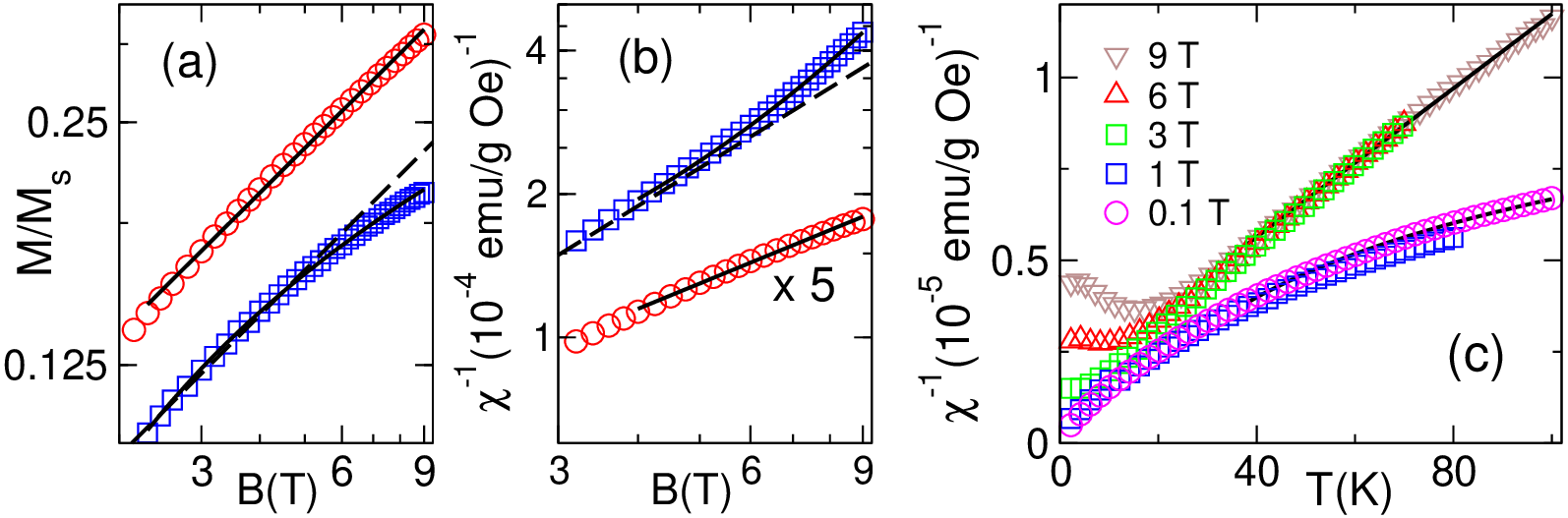}
\caption{(a) Magnetization normalized by the saturation magnetization M$_s$=$g\mu_BNs$
and (b) a.c magnetic susceptibility as a function of the applied magnetic
field for $\mbox{MgTiOBO$_3$}$ (open squares) and $\mbox{MgMnB$_2$O$_5$}$
(open circles). The solid lines represent a power law fitting for the $ 
\mbox{MgMnB$_2$O$_5$}$ data and a logarithmic fitting for the $ 
\mbox{MgTiOBO$_3$}$ data. c) a.c magnetic susceptibility as a function of
temperature for different values of an applied d.c field ($\mbox{MgTiOBO$_3$}
$). The solid line indicates a Curie-law fitting at high temperatures.}
\label{Fig2}
\end{figure}

We further investigate these two systems by comparing other independent
measurements, such as magnetization and ac susceptibility as a function of a
magnetic field, as shown in in Fig.~\ref{Fig2}. From Fig.~\ref{Fig2}(a) and
Fig.~\ref{Fig2}(b) we see that the $\mbox{MgTiOBO$_3$}$\ data always present
logarithmic corrections and the magnetization follows $M\propto \ln (B)$, as
expected for a RSP~\cite{fisher94}. On the other hand, for $%
\mbox{MgMnB$_2$O$_5$}$\ both $\chi (B)$ and $M(B)$ follow a power law
behavior with exponents $\alpha \sim 0.55$ and $1-\alpha \sim 0.45$
respectively. Such behavior is expected for systems in a GP. Finally, for
the $\mbox{MgTiOBO$_3$}$, we also analyze $\chi _{\mbox{a.c}}\times T$ for
different applied fields $B$ (Fig.~\ref{Fig2}(c) ): the RSP is robust to
applied fields bellow 3T even at temperatures up to 100K, where the
susceptibility is not Curie-like. However, the RSP characteristics disappear
at high temperatures once the applied field is around 3T with the appearance
of a Curie-like behavior.

%%%%%%%%%%%%%%%%%%%%%%%%%%%%

%%%%%%%%%%%%%%%%%%%%%%%%%
Once established that $\mbox{MgTiOBO$_3$}$\ is in a RSP, we can expect the
system to be entangled. Theoretically, the entanglement can be estimated by
calculating the Von Neumann entanglement entropy of a subsystem A of the
spin chain, with respect to a subsystem B. This quantity can be defined as $%
S=-\mbox{Tr}\hat \rho_A\ln \hat \rho_A$. For a subsystem with length $x$
embedded in an infinite system, the entanglement entropy for a Heisenberg
chain in a RSP is given by $S=\frac{\ln(2)}{3}\ln(x)$, as previously
calculated by means of a RSRG~\cite{refael04} and further confirmed by
numerical calculations~\cite{laflorencie}. From an experimental point of
view, it is not possible to use the entanglement entropy to quantify the
entanglement. Entanglement witnesses (EW) have been proposed and applied as
an attempt to detect and quantify entanglement experimentally. The main
advantage of EW is that they are observables and their mean value can be
directly measured.

The use of magnetic susceptibility as an EW is based on the violation of
local uncertainty relations~\cite{hofmann}. The uncertainty of an operator $%
\hat{A}_i$ for a given quantum state is defined as $\Delta A_i^2=\langle
A_i^2\rangle - \langle A_i\rangle^2$, the statistical variance of the
measurement outcomes . The uncertainty $\Delta A^2$ can only be zero if the
quantum state is an eigenstate of $\hat{A_i}$. A quantum state with zero
uncertainty in all the properties $\hat{A_i}$, must be a simultaneous
eigenstate of these operators. If the simultaneous eigenstate does not
exist, there must be a lower limit for the sum of the uncertainties. We can
illustrate this concept for spins where $s=(N-1)/2$: for any given spin
state, we have $(\hat{S_x}^2+\hat{S_y}^2+\hat{S_z}^2)|\psi\rangle=s(s+1)|%
\psi\rangle$. On the other hand, the expectation values of $\hat{S_i}$
defines a vector with maximal length equal to $\pm s$. Using both relations,
we obtain the inequality $\Delta S_{x}^{2}+\Delta S_{y}^{2}+\Delta S_{z}^{2}
=\langle \hat{S_{x}} ^{2}+\hat{S_{y}}^{2}+\hat{S_{z}}^{2}\rangle -(\langle 
\hat{S_{x}}\rangle ^{2}+\langle \hat{S_{y}}\rangle ^{2}+\langle \hat{S_{z}}%
\rangle ^{2}) \geq s$

We can apply this relation to obtain a limit for the correlation of
separable states. Exemplifying for two spins 1 and 2: for product states,
the uncertainty of $\hat{S_i}(1) + \hat{S_i}(2)$ is equal to the sum of the
local uncertainties. On the other hand, maximally entangled states can have
a total uncertainty of zero in all directions of $\hat{S_i}(1) + \hat{S_i}%
(2) $, \textit{maximally violating the uncertainty relation}. In this case,
the quantity $E=1-\sum_i\frac{\Delta(\hat{S_i}(1) + \hat{S_i}(2))^2}{2s}$
measures the amount of entanglement verified by the violation of local
uncertainties~\cite{hofmann,wquant}. For a macroscopic system, we can
generalize this quantity by using the magnetic susceptibility, which can be
written as $\chi_i=\frac{1}{k_BT}\Delta^2 M_i =\frac{1}{k_BT}(\langle
M_i^2\rangle -\langle M_i \rangle^2)$.

Following ref.\onlinecite{vedral}, the entanglement can be measured by the
quantity 
\begin{equation}
E=1-k_{B}T\left( \frac{\chi _{x}+\chi _{y}+\chi _{z}}{(g\mu _{B})^{2}Ns}%
\right) .  \label{entan}
\end{equation}%
In our case, $N$ is the total number of spins per gram. First, we analyze a
specific limit: if there is no dc applied field and the system is isotropic, 
$E=1-3k_{B}T\chi _{z}/(g^{2}{\mu _{B}}^{2}Ns)$. Since the studied samples
have vanishing magnetic anisotropy, only one component of the susceptibility
is needed to detect and quantify entanglement. For $\chi _{z}<\frac{g^{2}\mu_B
^{2}Ns}{3k_{b}T}$, the system is entangled, and $E$ quantifies the
entanglement verified by the EW.

%%%%%%%%%%%%%%%%%%%%%%%%%%%%%%%%%%%%%%%%%%%%%%%%%%%%%%%%%%%%
\begin{figure}[t]
\includegraphics[width=0.8\columnwidth,clip]{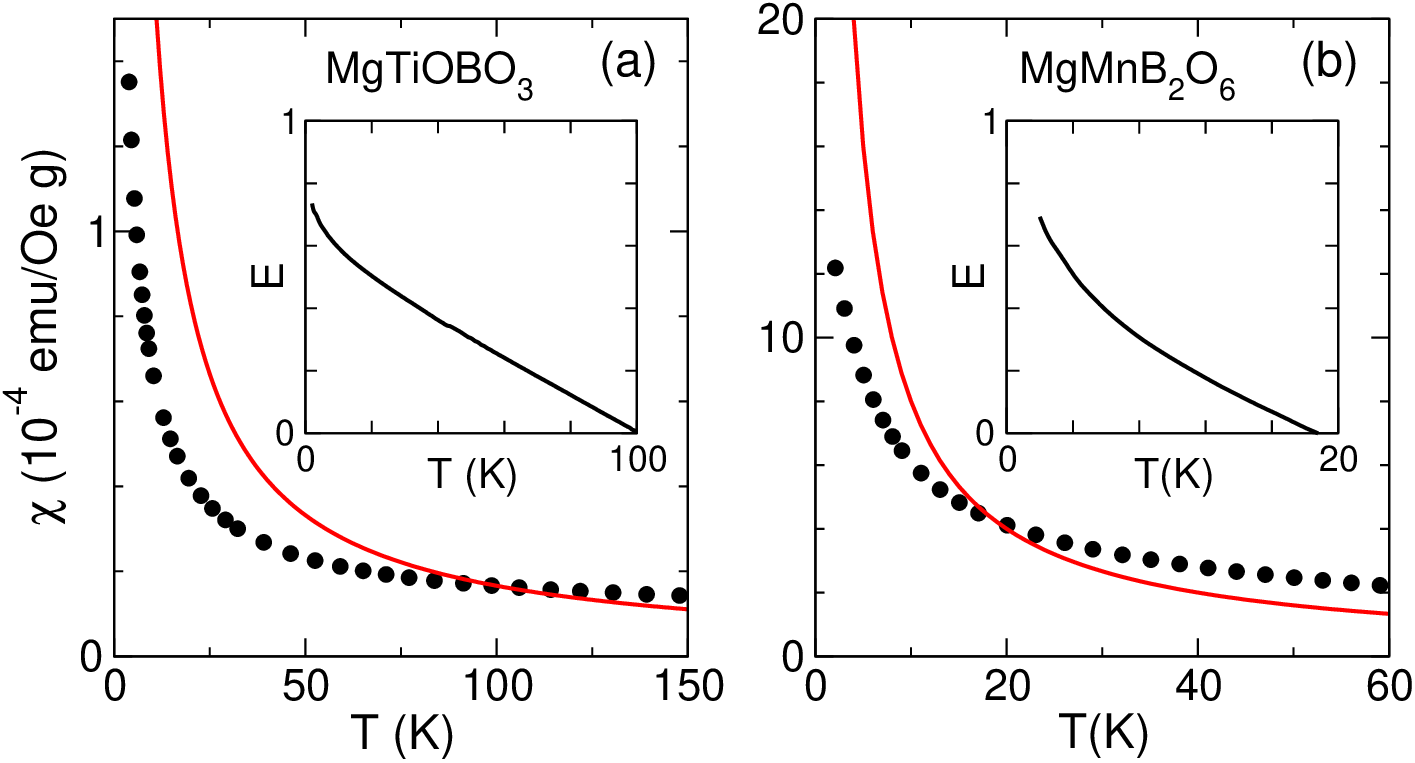}
\caption{(a) Experimental data of magnetic susceptibility for $ 
\mbox{MgTiOBO$_3$}$\ (closed circles) and the limit for the EW (solid line).
The system presents entanglement below the solid line. On panel (b) we show
the same analysis for $\mbox{MgMnB$_2$O$_5$}$. The insets show the quantity
E as defined in eq. \protect\ref{entan} which quantifies the entanglement
detected by the EW as a function of temperature. }
\label{Fig3}
\end{figure}
In Fig.~\ref{Fig3} we show the experimental data for $\mbox{MgTiOBO$_3$}$
and $\mbox{MgMnB$_2$O$_5$}$: both systems present entanglement, although $%
\mbox{MgTiOBO$_3$}$\ is entangled up to higher temperatures. The quantity $E$
has a similar behavior as a function of temperature for both compounds, with
a linear dependence on $T$ for high temperatures.

\begin{figure}[t]
\center{
\includegraphics[width=0.9\columnwidth,clip]{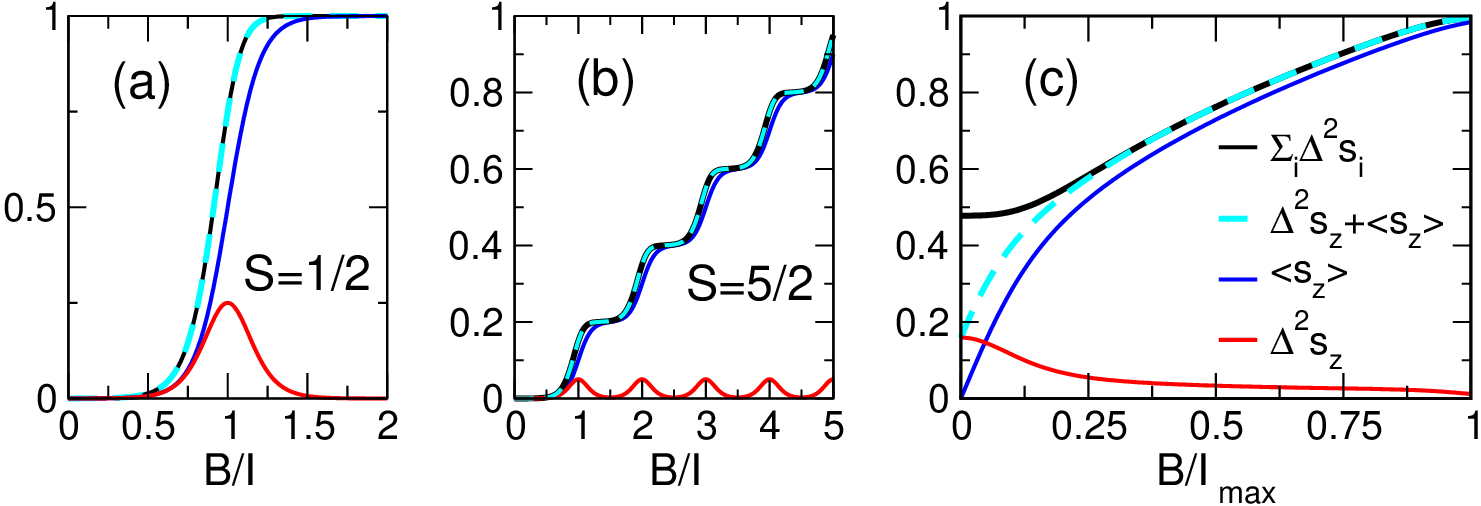}
\includegraphics[width=0.9\columnwidth,clip]{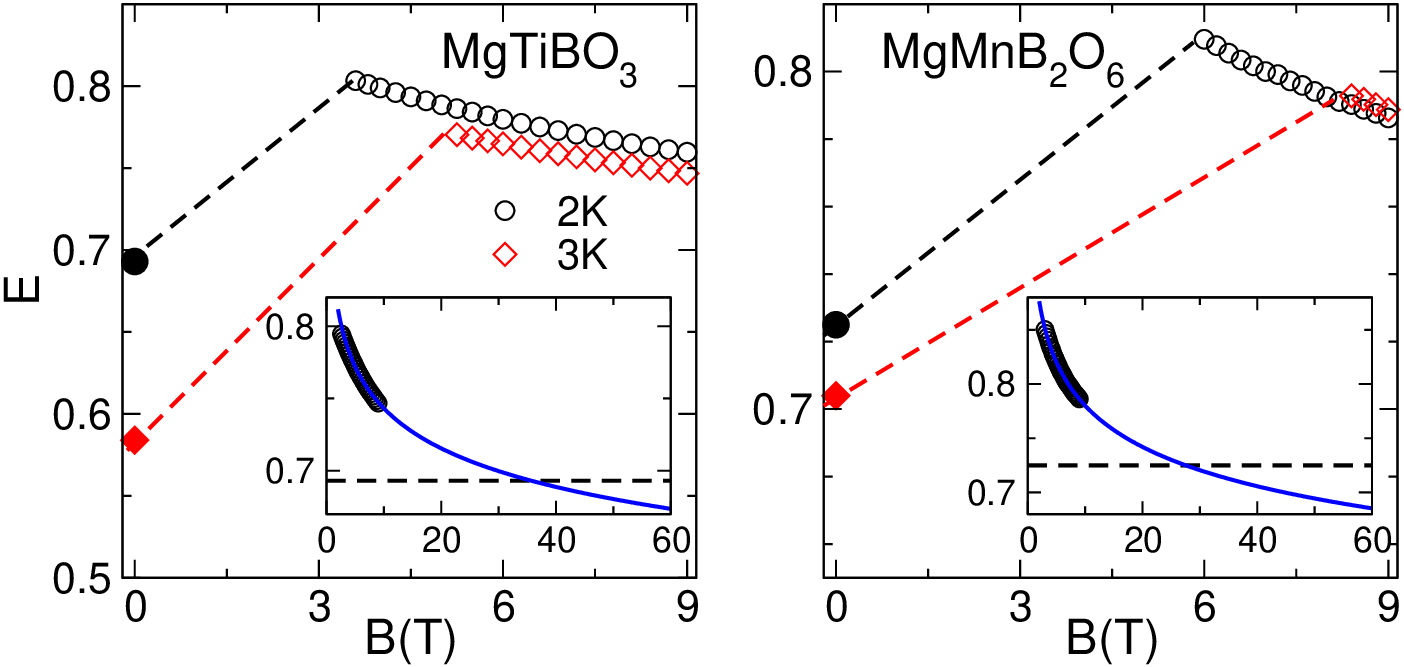}}
\caption{Up: Expectation value of the z component of the total spin $\langle
J_{z}\rangle $, the sum of the total spin variance in three directions $%
\sum_{i}\Delta ^{2}J_{i}$, the spin variance in the z direction $\Delta
^{2}J_{z}$ and the sum $\Delta ^{2}J_{z}+\langle J_{z}\rangle $ for (a) a
pair of spins S=1/2, (b) a pair of spin S=5/2 and (c) pairs of spin S=1/2
interaction with a random interaction $I_{i}$ (power-law distribution and
T=0.05) as a function of magnetic field. This is normalized by
the exchange interaction and in the random case by the cutoff of the
distribution.. Down: Experimental data for E using eq. \protect\ref{entan2} for $%
B\neq 0$ for $\mbox{MgTiOBO$_3$}$\ and $\mbox{MgMnB$_2$O$_5$}$. The insets
show the extrapolation of $E$ for very high values of $B$.}
\label{Fig4}
\end{figure}

For an applied dc field in the z direction, a pair of spins 1/2, where $\hat{%
J}=\hat{S_{1}}+\hat{S_{2}}$, form a singlet ($\mathcal{H}=I\hat{S_{1}}\cdot 
\hat{S_{2}}$) at low fields. As $[\mathcal{H},B\hat{J_{z}}]=0$, the field
does not modify the eigenstates, changing only their energies. At a given
field, the energy of the singlet is no longer the lowest energy $B_{c}$, the
singlet breaks and the spins align with the field $B$. However, for the
whole range of fields, the ground state of the system is such that spin
variance is minimal: $\Delta J_{x}\Delta J_{y}=\frac{1}{2}\langle
J_{z}\rangle $. In this case, as the system is isotropic in the xy plane, we
have $\Delta ^{2}J_{y}=\Delta ^{2}J_{x}=\frac{1}{2}\langle J_{z}\rangle $ so 
$\Delta ^{2}J_{y}+\Delta ^{2}J_{x}=\langle J_{z}\rangle $. This
approximation is valid for $g\mu _{B}sB\ll k_{B}T$, which assures that other
states, which do not have this property, are not populated. Similarly, the
same approach holds for two pairs of spin 5/2 as shown in Figs. \ref{Fig4}%
(a) and \ref{Fig4}(b). As an illustration, we also consider a distribution
of singlets, where the probability for interaction strength $I$ follows a
power law. As can be seen in panel (c) of the same figure, the approximation
works well for high values of the magnetic field compared with the
temperatures. Since both systems are in a phase where the spins form dimers (%
$\mbox{MgTiOBO$_3$}$ is in a RSP and $\mbox{MgMnB$_2$O$_5$}$\ presents
Griffiths singularities in a random dimer phase) we can use this
approximation to study quantum chains. It is possible to re-write the EW as 
\begin{equation}
E=1-\left( \frac{M_{z}}{g\mu _{B}Ns}+k_{B}T\frac{\chi _{z}}{(g\mu _{B})^{2}Ns%
}\right) ,  \label{entan2}
\end{equation}%
which is valid only at high fields. We perform the necessary measurements
and using Eq.~\ref{entan} for $B=0$ and Eq.~\ref{entan2} for high magnetic
fields ($g\mu _{B}sB>6k_{B}T$ for $\mbox{MgMnB$_2$O$_5$}$ and $g\mu
_{B}sB>2k_{B}T$ for $\mbox{MgTiOBO$_3$}$) we quantify the entanglement for
both systems. In Fig. \ref{Fig4} we unambiguously show that the magnetic
field can increase entanglement in quantum spin chains, as theoretically
suggested {vedral1, saguia1}. In the insets, we extrapolate the behavior of $%
E\times B$ for higher field values; measurements were performed with fields
up to 9 T. From this extrapolation, we see that even if a field of 9T is not
high enough for the approximation made in eq \ref{entan2}, the extrapolation
shows that the suggested increase of entanglement for low fields is still
valid although the amount could be slightly overestimated. 
%%%%%%%%%%%%%%%%%%%%%%%%%%%%%%%%

In conclusion, by means of macroscopic magnetic measurements we
fully characterize a random singlet phase in a low dimensional spin system
and for the first time, it was possible to distinguish this phase from a
Griffiths phase. We use a novel analysis where the magnetic susceptibility 
plays the role of an entanglement witness and measure the entanglement in two
different spin systems as a function of temperature and magnetic field. We
believe that both types of analysis presented here can be used to experimentaly 
characterize the phase diagram of low dimensional systems.

T. G. R would like to thank the group of quantum optics at IF-UFRJ  and P. Milman for
useful discussions and KITP at UCSB for the hospitality. This work was
partially supported by CNPq,  FAPERJ and the NSF (grant No. PHY99-07949).

\end{document}